\documentclass[prl,twocolumn,showpacs,superscriptaddress]{revtex4}
\usepackage{graphicx}
\usepackage{latexsym}
\usepackage{amsmath}
\newcommand{\bq}{\begin{equation}}
\newcommand{\eq}{\end{equation}}
\newcommand{\bqn}{\begin{eqnarray}}
\newcommand{\eqn}{\end{eqnarray}}
\newcommand{\nb}{\nonumber}
\newcommand{\lb}{\label}
\begin{document}
\title{Friedmann Equations and Thermodynamics of Apparent Horizons}

\author{Yungui Gong}
\email[]{gongyg@cqupt.edu.cn}
\affiliation{College of Mathematics and Physics, Chongqing
University of Posts and Telecommunications, Chongqing 400065,
China}
\affiliation{GCAP-CASPER, Department of Physics, Baylor University,
Waco, TX 76798, USA}
\author{Anzhong Wang}
\email[]{anzhong_wang@baylor.edu}
\affiliation{GCAP-CASPER, Department of Physics, Baylor University,
Waco, TX 76798, USA}


\begin{abstract}
With the help of a masslike function which has dimension of energy and equals to
the Misner-Sharp mass at the apparent horizon, we show that the first law of
thermodynamics of the apparent horizon $dE=T_AdS_A$ can be derived from the
Friedmann equation in various theories of gravity, including the Einstein,
Lovelock, nonlinear, and scalar-tensor theories. This result strongly suggests
that the relationship between the first law of thermodynamics of the apparent
horizon and the Friedmann equation is not just a simple coincidence, but rather
a more profound physical connection.
\end{abstract}
\maketitle
The derivation of the thermodynamic laws of black holes from the classical
Einstein equation suggests a deep connection between gravitation and thermodynamics \cite{bardeen}.
The discovery of the quantum Hawking radiation \cite{hawking} and black hole
entropy which is proportional to the area of the event horizon of the black hole
\cite{bekenstein} further supports this connection and the thermodynamic (physical) interpretation
of geometric quantities.
The interesting relation between thermodynamics
and gravitation became manifest when Jacobson derived Einstein equation
from the first law of thermodynamics by assuming
the proportionality of the entropy and the horizon area
for all local acceleration horizons \cite{ted}.

In cosmology, like in black holes, for the cosmological model
with a cosmological constant (called de Sitter space), there
also exist Hawking temperature and entropy associated with the
cosmological event horizon, and thermodynamic laws of the
cosmological event horizon \cite{gibbons}. In de Sitter space, the event horizon coincides
with the apparent horizon (AH). For more general cosmological models, the event horizon
may not exist, but the AH always exists, so it is possible to have
Hawking temperature and entropy associated with the AH. The connection
between the first law of thermodynamics of the AH and the Friedmann equation
was shown in \cite{cai05}. Now, we must ask if this interesting relation between gravitation
and thermodynamics exists in more general theories of gravity, like Brans-Dicke (BD) theory and
nonlinear gravitational theory. In \cite{eling}, the gravitational field equations for
the nonlinear theory of gravity were
derived from the first law of thermodynamics
by adding some nonequilibrium corrections.
In this Letter, we show that equilibrium thermodynamics indeed exists for more general
theories of gravity, provided that a new masslike function is introduced.

To show our claim, we begin by
reviewing the thermodynamics of the AH with the use of the Misner-Sharp (MS) mass in
Einstein and BD theories of gravity, whereby we find the
equilibrium thermodynamics fails to hold for the BD theory.
The Einstein equation can be rewritten as the mass formulas with the help of the MS
mass $\mathcal{M}$. The energy flow through the AH $dE$ is related with the MS
mass. Since the MS mass $\cal{M}$, the Hawking temperature $T_A$, and the entropy
$S_A$ of the AH are geometric quantities, the first law of thermodynamics of
the AH can be thought of as a geometric relation. Therefore, we expect
the geometric relation to hold in other gravitational theory if it holds in Einstein theory.
To achieve this, we replace the MS mass $\cal{M}$ by a masslike function
$M$ which equals to the MS mass $\cal{M}$ at the AH, we then
show that the connection between the first law of thermodynamics of the AH
and the gravitational equations holds in scalar-tensor and nonlinear
theories of gravity without adding nonequilibrium correction.

For a spherically symmetric space-time
with the metric $ds^2=g_{ab}dx^a dx^b+\tilde{r}^2 d\Omega^2$,
using the MS mass
${\cal{M}}=\tilde{r}(1 - g^{ab}\tilde{r}_{,a}\tilde{r}_{,b})/2G$ \cite{misner},
the $a-b$ components of the Einstein equation give the mass formulas \cite{israel,gong}
\begin{equation}
\lb{msmasseq}
{\cal{M}}_{,a}=4\pi\tilde{r}^2(T_a^b-\delta_a^b T)\tilde{r}_{,b},
\end{equation}
where the unit spherical metric is given by $d\Omega^2=d\theta^2+\sin^2\theta d\varphi^2$
and $T=T^a_a$. From now on, all the indices are raised and lowered by the metric $g_{ab}$
and the covariant derivative is with respect to  $g_{ab}$.
The AH is
\bq
\lb{apphor}
\tilde{r}_A=ar_A=(H^2+k/a^2)^{-1/2}.
\eq
At the AH, the MS mass ${\cal M}=4\pi\tilde{r}_A^3\rho/3$, which
can be interpreted as
the total energy inside the AH. Now we use the (approximate)
generator $k^a=(1,\ -Hr)$ of the
AH, which is null at the horizon, to project the mass formulas.
Since $k^a\tilde{r}_{,a}=0$, at the AH we find that
\bq
\lb{energyeq}
-dE=-k^a\nabla_a {\cal M} dt=d(\tilde{r}_A)/G=T_A dS_A,
\eq
where the horizon temperature is $T_A=1/(2\pi \tilde{r}_A)$
and the horizon entropy is $S_A=\pi\tilde{r}_A^{2}/G$.
 On the other hand, using
the mass formulae (\ref{msmasseq}), we get the energy flow through the AH
\bqn
\lb{energyflux}
-dE &=& -k^a\nabla_a {\cal M} dt=-4\pi \tilde{r}^2 T_a^b\tilde{r}_{,b}k^adt\nb\\
&=& 4\pi\tilde{r}_A^3H(\rho+p)dt.
\eqn
Therefore, the Friedmann equation gives rise to the first law of
thermodynamics $-dE=T_A dS_A$ of the
AH. From the above definitions, we see that
the relation $-dE=T_A dS_A$ is a geometrical relation which depends on
the only assumption of the Robertson-Walker metric. To connect the geometrical quantity $dE$
with the energy flow through the AH, we need to use the Friedmann
equations. Therefore, for any gravitational theory, if we can write the gravitational field equation as
$G_{\mu\nu}=8\pi G T_{\mu\nu}$ and regard the right-hand side as the effective
energy-momentum tensor, then we find the energy flow through the AH,
whereby we derive the first law of thermodynamics
of the AH $-dE=T_A dS_A$. For example, in the Jordan frame of the
scalar-tensor theory of gravity, if we take the right-hand side of gravitational field equation
as the total effective energy-momentum tensor, then the Friedmann equation
can be regarded as a thermodynamic identity at the AH \cite{cai06}.

The connection between the first law of thermodynamics and
the Friedmann equation at the AH was also found for gravity with Gauss-Bonnet term,
the Lovelock  theory of gravity \cite{cai05}, and the braneworld cosmology
\cite{ge}. For a general static spherically symmetric and stationary axisymmetric space-times,
it was shown that Einstein equation at the horizon give rise to
the first law of thermodynamics \cite{padmanabhan02,dawood}. For the Lovelock gravity, the interpretation
of gravitational field equation as a thermodynamic identity was proposed in
\cite{padmanabhan06}.

Alternatively, the mass formulae (\ref{msmasseq}) can be written as the so-called unified first law
$\nabla_a {\cal{M}} =A\Psi_a+W\nabla_a  V$ \cite{sean,sean1}, where $W=(\rho-p)/2$
and $\Psi_a=T_a^b\tilde{r}_{,b}+W\tilde{r}_{,a}$. Projecting the
unified first law along the direction tangent to the AH
(or trapping horizon in Hayward's terminology), the first law of thermodynamics
$d{\cal{M}} =TdS+WdV$ can be
derived, where the horizon temperature and entropy are given, respectively,
by $T=\Box \tilde{r}/(4\pi)$ and $S=A/(4G)$. Based on this result,
the connection between the Friedmann equation and the first law of
thermodynamics of the AH with the work term was widely discussed for Einstein
gravity, Lovelock's gravity, the scalar-tensor theory of gravity,
the nonlinear theory of gravity, and the braneworld scenario
\cite{akbar,rgcao,rgcao1,akbar1,wang,wang1}.

This connection between the Friedmann equation and the first law of thermodynamics
of the AH suggests the unique role of the AH in thermodynamics
of cosmology. This may be used to probe the property of dark energy \cite{gong,gong1}.
For example, if we assume that the temperature of the dark components is $T=bT_A$, then use the relation
$T=(\rho+p)/s=(\rho+p)a^3/\sigma$, we find that the total energy density
of the dark components is given by
\bq
\lb{dedens}
\rho=\rho_\Lambda+\rho_0\left(\frac{a_0}{a}\right)^6
+2\sqrt{\rho_0\rho_\Lambda}\left(\frac{a_0}{a}\right)^3,
\eq
where $\rho_0=\sigma^2b^2 Ga^{-6}_0/(6\pi)$, $\rho_\Lambda=3\Lambda/(8\pi G)$ is the
energy density of the cosmological constant, $\sigma$ is the constant comoving entropy
density, and $s$ is the physical entropy density. The right-hand side of the above
equation contains three different terms, which correspond to,  respectively,
the cosmological constant, the stiff fluid, and the pressureless matter. However,
the coefficients of these terms are not all independent. In fact, the current
observational constraints tell us that the stiff fluid is negligibly small,
for which we must assume $\rho_{0} \ll 1$. This  in turn implies that the pressureless
matter given by the last term is also negligibly small. So the pressureless
matter in the last term cannot account for dark matter. In other words,
the dark matter must not be in equilibrium with the AH.

For the BD theory \cite{bd} \bq \lb{bdaction}
L=-\frac{\sqrt{-g}}{16\pi}\left[\phi R+\omega g^{\mu\nu}
\frac{\partial_\mu\phi\partial_\nu\phi}{\phi}\right], \eq the
BD scalar $\phi$ plays the role of the gravitational
constant. The MS mass is \cite{bdms}
\bq
\lb{bdmsmass}
{\cal{M}}=\phi\tilde{r}(1 -
g^{ab}\tilde{r}_{,a}\tilde{r}_{,b})/2.
\eq
At the AH,
${\cal M}=\phi\tilde{r}_A/2$. The horizon entropy is
$S_A=\pi\tilde{r}^2_A\phi$, so we get
\bq
\lb{bdtds}
T_AdS_A=\tilde{r}_Ad\phi/2+\phi d\tilde{r}_A.
\eq
On the other hand, we have
\bq
\lb{bdde}
-dE=-k^a\nabla_a{\cal M}
dt=-\tilde{r}_Ad\phi/2+\phi d\tilde{r}_A.
\eq
Comparing Eqs.\,(\ref{bdtds}) with (\ref{bdde}), we find that the equilibrium
thermodynamics $-dE=T_A dS_A$ fails to hold for the BD
theory. Similarly, it can be shown that $-dE=T_A dS_A$ does not hold
in the nonlinear and scalar-tensor theories of gravity. It is
exactly because of this that it  was argued nonequilibrium
treatment might be needed.

As mentioned above,
the  mass, temperature  and  entropy of the AH are all geometrical
quantities, and the first law of thermodynamics
of the AH can be regarded as a geometric relation.
Now, the important question is whether a mass function exists that serves
as the bridge between the Friedmann equation and the first law of thermodynamics
of the AH without nonequilibrium correction.
In the following, we show that the answer is affirmative.
It has exactly the dimension of energy, and is equal to the MS mass
at the AH. To distinguish it with the MS mass,
we call it the masslike function.

To show our above claim, let us write
the $a-b$ components of the Einstein equation as
\begin{equation}
\lb{masseq}
M_{,a}=-4\pi\tilde{r}^2(T_a^b-\delta_a^b T)\tilde{r}_{,b}+\tilde{r}_{,a},
\end{equation}
where the mass-like function $M$ is defined as
\bq
\lb{mfunc}
M\equiv \frac{\tilde{r}}{2G}(1+g^{ab}\tilde{r}_{,a}\tilde{r}_{,b}).
\eq
At the AH, $g^{ab}\tilde{r}_{,a}\tilde{r}_{,b}=0$ and
the masslike function $M=\tilde{r}_A/2G$, which is equal to the MS mass.
For the Robertson-Walker metric we have $g_{tt}=-1$,
$g_{rr}=a^2/(1-kr^2)$ and $\tilde{r}=ar$.
Then, the mass formulas (\ref{masseq})
yield the Friedmann equations
\bq
\lb{frweq1}
H^2+\frac{k}{a^2}=\frac{8\pi G}{3}\rho,
\eq
\bq
\lb{frweq2}
\frac{\ddot{a}}{a}=-\frac{4\pi G}{3}(\rho+3p).
\eq
Combining Eqs.\,(\ref{frweq1}) and (\ref{frweq2}), we can derive
the energy conservation law $\dot{\rho}+3H(\rho+p)=0$. Thus,
the mass formulas (\ref{masseq}) give rise to the full set of the
cosmological equations.

At the AH, the masslike function $M=4\pi\tilde{r}^3_A\rho/3$, which
is the total energy inside the AH. The energy flow is
\bq
\lb{energyeq1}
dE=k^a\nabla_a M dt=d(\tilde{r}_A)/G=T_A dS_A.
\eq
On the other hand, using
the mass formulas (\ref{masseq}), we get the energy flow through the AH
\bqn
\lb{energyflux1}
dE &=& k^a\nabla_a M dt=-4\pi \tilde{r}^2 T_a^b\tilde{r}_{,b}k^adt\nb\\
&=& 4\pi\tilde{r}_A^3H(\rho+p)dt.
\eqn
Therefore, the Friedmann equation gives rise to the first law of
thermodynamics $dE=T_A dS_A$ of the AH. While this result is the same as
that obtained by using the MS mass, we show below that the equilibrium
thermodynamics can be derived for BD and nonlinear gravities by using our
newly defined masslike function, although it cannot be done by using
the MS mass, as shown above.

For the BD theory, the mass-like function is defined as
\bq
\lb{bdmass}
M \equiv \phi\tilde{r}(1+g^{ab}\tilde{r}_{,a}\tilde{r}_{,b})/2.
\eq
At the AH, it reduces to the MS mass,
$M={\cal M}=\phi\tilde{r}_A/2$. The $a-b$ components
of the gravitational field equation become
\bq
\lb{bdmeq}
\begin{split}
M_{,a}=&-4\pi\tilde{r}^2(T_a^b-\delta^b_a T)\tilde{r}_{,b}
+2\pi\tilde{r}^3 T\frac{{\phi}_{,a}}{\phi}+(\phi\tilde{r})_{,a}\\
&-\frac{\omega+2}{2\phi}\tilde{r}^2\phi_{,a}\phi_{,b}\tilde{r}^{;b}
+\frac{\omega}{4\phi}\tilde{r}^2\tilde{r}_{,a}\phi_{,b}\phi^{;b}
-\tilde{r}\tilde{r}_{,a}\tilde{r}_{,b}\phi^{;b}\\
&-\frac{1}{2}\tilde{r}^2
\phi_{;ab}\tilde{r}^{;b}-\frac{1}{2}\tilde{r}^2\Box\tilde{r}\phi_{,a}
-\frac{\tilde{r}^3}{4\phi}\phi_{,a}\Box\phi.
\end{split}
\eq
Applying to the Robertson-Walker metric, the above equation gives the
Friedmann equations
\bq
\lb{bdfrw1}
H^2+\frac{k}{a^2}=\frac{8\pi}{3\phi}\rho+\frac{\omega}{6}\frac{\dot{\phi}^2}{\phi^2}
-H\frac{\dot{\phi}}{\phi},
\eq
\bq
\lb{bdfrw2}
\frac{\ddot{a}}{a}=-\frac{4\pi}{3\phi}(\rho+3p)-\frac{\omega}{3}\frac{\dot{\phi}^2}{\phi^2}
-\frac{1}{2}H\frac{\dot{\phi}}{\phi}-\frac{1}{2}\frac{\ddot{\phi}}{\phi}.
\eq
The mass formulas (\ref{bdmeq}) or Eqs.\, (\ref{bdfrw1}) and (\ref{bdfrw2}) are
not sufficient to describe the full
dynamics of the BD cosmology. In the BD cosmology, we also need
the equation of motion of the BD scalar field $\phi$
in addition to   Eqs. (\ref{bdfrw1}) and (\ref{bdfrw2}), which is given by
\bq
\lb{bdphi}
\ddot{\phi}+3H\dot{\phi}=\frac{8\pi}{3+2\omega}(\rho-3p).
\eq
From the definition of the masslike function (\ref{bdmass}), at the AH we find
\bq
\lb{bddm}
dE=M_{,a}k^adt=\tilde{r}_A d\phi/2+\phi d\tilde{r}_A=T_A dS_A,
\eq
where the entropy now is $S_A=\pi\tilde{r}^2_A\phi$. Using the mass formulas (\ref{bdmeq}),
we get the energy flow through the AH
\bq
\lb{bdenergy}
\begin{split}
M_{,a}k^a
=&\frac{8\pi}{3+2\omega}\tilde{r}_A^3H[(\omega+2)\rho+\omega p]+
\frac{\omega}{2}\tilde{r}_A^3H\frac{\dot{\phi}^2}{\phi}\\
&-2\tilde{r}_A^3 H^2\dot{\phi}+\frac{1}{2}\tilde{r}_A\dot{\phi},
\end{split}
\eq
where we used Eq. (\ref{bdphi}) in deriving the above equation.
From Eqs.\,(\ref{bdfrw1})-(\ref{bdphi}), the right-hand side of
Eq.\,(\ref{bdenergy}) can be written as
 $\frac{1}{2}\tilde{r}_A\dot{\phi}+\phi\dot{\tilde{r}}_A$.
Therefore, we see that in BD theory, the first law of thermodynamics of
the AH $dE=T_AdS_A$ can be derived from the Friedmann equation.

The thermodynamic prescription can be easily extended to
general scalar-tensor theory of gravity with the Lagrangian
\bq
\lb{ST}
L=f(\phi)R-  g^{\mu\nu}\partial_\mu\phi
 \partial_\nu\phi/2 -V(\phi).
\eq
In this case, $f(\phi)$ plays the role of
the gravitational constant, so now we can define the mass-like function as
\bq
\lb{STmass}
M \equiv f(\phi)\tilde{r}
\left(1+g^{ab}\tilde{r}_{,a}\tilde{r}_{,b}\right)/2,
\eq
and the horizon entropy as $S_A=\pi\tilde{r}^2_A f(\phi)$. Then, using these
definitions, we can show that $dE=M_{,a} k^a dt=T_A dS_A$.

For the nonlinear theory of gravity $f(R)$, we can define the masslike function as
\bq
\lb{STmass2}
M\equiv \frac{1}{2}f'(R)\tilde{r}\left(1+g^{ab}\tilde{r}_{,a}\tilde{r}_{,b}\right),
\eq
and the horizon entropy $S_A=\pi\tilde{r}^2_A f'(R)$, where $f'(R)=df/dR$.
Again, it is easy to show that $dE=M_{,a} k^a dt=T_A dS_A$. Therefore,
the thermodynamics of the AH holds for both the general
scalar-tensor theory of gravity and the nonlinear theory of gravity.

Now we show how to derive the first law of thermodynamics of the AH
from the Friedmann equation in the Lovelock gravity. The Lovelock
Lagrangian is $L=\sum_{n=0}^{m} c_n L_n$ \cite{lovelock}, where
$$L_n=2^{-n}\delta^{\mu_1\nu_1\cdots \mu_n\nu_n}_{\alpha_1\beta_1\cdots\alpha_n\beta_n}
R^{\alpha_1\beta_1}_{\mu_1\nu_1}\cdots R^{\alpha_n\beta_n}_{\mu_n\nu_n}.$$
Using the Robertson-Walker metric, we obtain the Friedmann equations in $N+1$ dimensional
space-time
\bq
\lb{llfrw1}
\sum_{i=1}^m \hat{c}_i \left(H^2+\frac{k}{a^2}\right)^i=\frac{16\pi G}{N(N-1)}\rho,
\eq
and
\bq
\lb{llfrw2}
\sum_{i=1}^m  \hat{c}_i i \left(H^2+\frac{k}{a^2}\right)^{i-1}\left(\dot{H}-\frac{k}{a^2}\right)
=-\frac{8\pi G}{N-1}(\rho+p),
\eq
where $\hat{c}_0=c_0/[N(N-1)]$, $\hat{c}_1=1$ and $\hat{c}_i=c_i\prod_{j=3}^{2m}(N+1-j)$ for
$i>1$.
The masslike function can now be defined as
\bqn
\lb{llmass}
M&\equiv &\frac{N(N-1)\Omega_N\tilde{r}^N}{16\pi G}\sum_{i=1}^m\hat{c}_i
\left[2\tilde{r}^{-2i}-\left(H^2+\frac{k}{a^2}\right)^i\right] \nb\\
&=& \Omega_N\tilde{r}_A^N\rho,
\eqn
where $\Omega_N$ is the volume of unit $N$-dimensional sphere
and the last equality is evaluated at the AH. Note that although the geometric
form is different, the masslike function at the AH
has the same value as that in Einstein theory of gravity,
which is the total energy inside the AH. The entropy of
the AH is
\bq
\lb{lls}
S_A=\frac{N\Omega_N}{4G}\sum_{i=1}^m \frac{i(N-1)}{N-2i+1}\hat{c}_i\tilde{r}_A^{N+1-2i}.
\eq
From Eqs. (\ref{llmass}) and ({\ref{lls}), we can easily check that $dE=M_{,a} k^a dt=T_A dS_A$
holds with the horizon temperature $T_A=1/(2\pi\tilde{r}_A)$. Using the Friedmann Eqs.
(\ref{llfrw1}) and (\ref{llfrw2}), we find the energy flow through the AH
is $dE=N\Omega_N H\tilde{r}_A^N (\rho+p)$, which is the same as that in Einstein's gravity.

By properly
defining the masslike function in each theory of gravity, we find
that the corresponding Friedmann equations can be written in the
form $dE=T_A dS_A$ of the first law of thermodynamics at the
AH. In other words, the thermodynamic description of
the gravitational dynamics is manifest through the mass formulas.
Therefore,  the gravitational dynamics can be considered as the
thermodynamic identity $dE=T_A dS_A$. This is true for a variety of
theories of gravity, including the Einstein, Lovelock, nonlinear,
and scalar-tensor theories. This non-trivial connection between the
thermodynamics of the AH and the Friedmann equation
may represent a generic connection, and it suggests the unique role
that the AH can play in the thermodynamics of
cosmology. Such a thermodynamic description of the AH can
also be used to probe other physical systems and properties, such as
the nature of dark energy and the thermodynamics of black holes in
each of these theories.

Finally, we would like to note that, although the newly defined masslike function
reduces to the  MS mass at the AH, the corresponding energy
flows passing through the horizon are different. This explains  why our
masslike function gives rise to the first law of
thermodynamics in various theories of gravity, while the MS mass does not.
Because of the masslike function, the energy momentum tensor includes the contribution
of gravitational fields such as BD scalars, or curvature scalars in nonlinear
theory of gravity, in addition to the matter fields. This treatment allows a reinterpretation
of the nonequilibrium correction introduced in \cite{eling}.
The studies of other properties of the newly-defined masslike function, including
the physical and geometrical difference between the MS mass and it  are
important and should be reported somewhere else.

\begin{acknowledgments}
Y.G. Gong is supported by NNSFC under Grants
No. 10447008 and 10605042, CMEC under Grant No. KJ060502, and
SRF for ROCS, State Education Ministry. A. Wang's work was partially supported
by a VPR fund from Baylor University.
\end{acknowledgments}

\end{document}